\documentclass[12pt]{article}
\newcommand{\sect}[1]{\setcounter{equation}{0}\section{#1}}

\renewcommand{\appendix}{\setcounter{section}{0}
\renewcommand{\thesection}{\Alph{section}}}

\def\g{\gamma}
\def\G{\Gamma}
\def\d{\delta}

\def\e{\epsilon}

\def\k{\kappa}
\def\l{\lambda}
\def\m{\mu}
\def\n{\nu}

\def\s{\sigma}
\def\S{\Sigma}

\def\be{\begin{equation}}
\def\ee{\end{equation}}
\def\ba{\begin{eqnarray}}
\def\ea{\end{eqnarray}}

\newcommand{\nn}{\nonumber\\}
\newcommand{\no}{\nonumber}
\begin{document}
\renewcommand{\thefootnote}{\fnsymbol{footnote}}

\newpage
\setcounter{page}{0}
\pagestyle{empty}

\begin{center}
{\Large{\bf Semi-simple extension of the (super) Poincar\'e algebra\\}}
\vspace{1cm}
{\bf Dmitrij V. Soroka\footnote{E-mail: dsoroka@kipt.kharkov.ua} and 
Vyacheslav A. Soroka\footnote{E-mail: vsoroka@kipt.kharkov.ua}}
\vspace{1cm}\\
{\it Kharkov Institute of Physics and Technology,\\
1, Akademicheskaya St., 61108 Kharkov, Ukraine}\\
\vspace{1.5cm}
\end{center}
\begin{abstract}
A semi-simple tensor extension of the Poincar\'e algebra is 
proposed for the arbitrary dimensions $D$. A supersymmetric also semi-simple
generalization of this extension is constructed in the $D=4$ dimensions. This
paper is dedicated to the memory of Anna Yakovlevna Gelyukh.



\end{abstract}

\newpage
\pagestyle{plain}
\renewcommand{\thefootnote}{\arabic{footnote}}
\setcounter{footnote}0

\sect{Introduction}

In the papers \cite{gios1,gios2,cj,ss1,dss0,dss} the Poincar\'e algebra for the generators of
the rotations $M_{ab}$ and translations $P_a$ in $D$ dimensions
\begin{eqnarray}
[M_{ab},M_{cd}]=(g_{ad}M_{bc}+g_{bc}M_{ad})-(c\leftrightarrow d),\nonumber
\end{eqnarray}
\begin{eqnarray}
[M_{ab},P_c]=g_{bc}P_a-g_{ac}P_b,\nonumber
\end{eqnarray}
\begin{eqnarray}\label{1.1}
[P_a,P_b]=0
\end{eqnarray}
has been extended by means of the second rank tensor generator $Z_{ab}$ in
the following way:
\begin{eqnarray}
[M_{ab},M_{cd}]=(g_{ad}M_{bc}+g_{bc}M_{ad})-(c\leftrightarrow d),\nonumber
\end{eqnarray}
\begin{eqnarray}
[M_{ab},P_c]=g_{bc}P_a-g_{ac}P_b,\nonumber
\end{eqnarray}
\begin{eqnarray}
[P_a,P_b]=cZ_{ab},\nonumber
\end{eqnarray}
\begin{eqnarray}
[M_{ab},Z_{cd}]=(g_{ad}Z_{bc}+g_{bc}Z_{ad})-(c\leftrightarrow d),\nonumber
\end{eqnarray}
\begin{eqnarray}
[P_a,Z_{bc}]=0,\nonumber
\end{eqnarray}
\begin{eqnarray}\label{1.2}
[Z_{ab},Z_{cd}]=0,
\end{eqnarray}
where $c$ is some constant\footnote{Note that, to avoid the double count under 
summation
over the pair antisymmetric indices, we adopt the rules which illustrated 
by the following example:
\ba
[P_a,P_b]=cZ_{ab}={c\over2}(\d_a^c\d_b^d-\d_a^d\d_b^c)Z_{cd}
=\sum_{c<d}{f_{ab}}^{cd}Z_{cd}={1\over2}{f_{ab}}^{cd}Z_{cd},\no
\ea
where ${f_{ab}}^{cd}$ are structure constants, and so on.}.

Such an extension makes common sense, since it is homomorphic to the usual 
Poincar\'e  algebra (\ref{1.1}). Moreover, in the limit ${c\to 0}$ the algebra
(\ref{1.2}) goes to the semi-direct sum of the commutative ideal $Z_{ab}$ and
Poincar\'e algebra (\ref{1.1}).

It is remarkable
enough that the momentum square Casimir operator of the Poincar\'e algebra
under this extension ceases to be the Casimir operator and it is generalized
by adding the term linearly dependent on the angular momentum
\begin{eqnarray}\label{1.3}
P^aP_a+cZ^{ab}M_{ba}\mathrel{\mathop=^{\rm def}}X_kh^{kl}X_l,
\end{eqnarray}
where $X_k=\{P_a, Z_{ab}, M_{ab}\}$. Due to this fact, an irreducible
representation of the extended algebra (\ref{1.2}) has to contain the fields 
of the different masses \cite{ss1,ss2}.
This extension with non-commuting momenta has also something in common
with the ideas of the papers~\cite{sn,ya,hl} and with the non-commutative
geometry idea~\cite{c}.

It is interesting to note that in spite of the fact that the algebra 
(\ref{1.2}) is not semi-simple and therefore has a degenerate Cartan-Killing
metric tensor nevertheless there exists another non-degenerate invariant
tensor $h_{kl}$ in adjoint representation which corresponds to the
quadratic Casimir operator (\ref{1.3}), where the matrix $h^{kl}$ is inverse
to the matrix $h_{kl}$: $h^{kl}h_{lm}=\d_m^k$.

There are other quadratic Casimir operators
\ba\label{1.4}
c^2Z^{ab}Z_{ab},
\ea
\ba\label{1.5}
c^2\e^{abcd}Z_{ab}Z_{cd}.
\ea
Note that the Casimir operator (\ref{1.5}), dependent on the Levi-Civita
tensor $\e^{abcd}$, is suitable only for the $D=4$ dimensions.

It has also been shown that for the dimensions $D=2,3,4$ the extended
Poincar\'e algebra (\ref{1.2}) allows the following supersymmetric
generalization:
\begin{eqnarray}
\{Q_\k,Q_\l\}=-d(\sigma^{ab}C)_{\k\l}Z_{ab},\nonumber
\end{eqnarray}
\begin{eqnarray}
[M_{ab},Q_\k]=-(\sigma_{ab}Q)_\k,\nonumber
\end{eqnarray}
\begin{eqnarray}
[P_a,Q_\k]=0,\nonumber
\end{eqnarray}
\begin{eqnarray}\label{1.6}
[Z_{ab},Q_\k]=0
\end{eqnarray}
with the help of the super-translation generators $Q_\k$. In (\ref{1.6}) $C$ 
is a charge conjugation matrix, $d$ is some constant and 
$\sigma_{ab}={1\over4}[\gamma_a,\gamma_b]$, where $\g_a$ is the Dirac matrix.
Under this supersymmetric generalization the quadratic Casimir operator
(\ref{1.3}) is modified into the following form:
\begin{eqnarray}\label{1.7}
P^aP_a+cZ^{ab}M_{ba}-{c\over2d}Q_\k(C^{-1})^{\k\l}Q_\l,
\end{eqnarray}
while the form of the rest quadratic Casimir operators (\ref{1.4}), 
(\ref{1.5}) remains unchanged.

In the present paper we propose another possible semi-simple tensor extension 
of the Poincar\'e algebra (\ref{1.1}) and for the case $D=4$ dimensions we 
give a supersymmetric generalization of this extension. In the limit this 
supersymmetrically generalized extension go to the Lie superalgebra
(\ref{1.2}), (\ref{1.6}).

\sect{Semi-simple tensor extension}

Let us extend the Poincar\'e algebra (\ref{1.1}) in the $D$ dimensions by 
means of the tensor generator $Z_{ab}$ in the following way:
\begin{eqnarray}
[M_{ab},M_{cd}]=(g_{ad}M_{bc}+g_{bc}M_{ad})-(c\leftrightarrow d),\nonumber
\end{eqnarray}
\begin{eqnarray}
[M_{ab},P_c]=g_{bc}P_a-g_{ac}P_b,\nonumber
\end{eqnarray}
\begin{eqnarray}
[P_a,P_b]=cZ_{ab},\nonumber
\end{eqnarray}
\begin{eqnarray}
[M_{ab},Z_{cd}]=(g_{ad}Z_{bc}+g_{bc}Z_{ad})-(c\leftrightarrow d),\nonumber
\end{eqnarray}
\begin{eqnarray}
[Z_{ab},P_c]={4a^2\over c}(g_{bc}P_a-g_{ac}P_b),\nonumber
\end{eqnarray}
\begin{eqnarray}\label{2.1}
[Z_{ab},Z_{cd}]={4a^2\over c}[(g_{ad}Z_{bc}+g_{bc}Z_{ad})
-(c\leftrightarrow d)],
\end{eqnarray}
where $a$ and $c$ are some constants. This Lie algebra, when the quantities
$P_a$ and $Z_{ab}$ are taken as the generators of a homomorphism kernel, is  
homomorphic to the usual Lorentz algebra.  It is remarkable that the Lie 
algebra (\ref{2.1}) is {\it semi-simple} in contrast to the Poincar\'e
algebra (\ref{1.1}) and extended Poincar\'e algebra (\ref{1.2}).

The extended Lie algebra (\ref{2.1}) has the following quadratic Casimir
operators:
\begin{eqnarray}\label{2.2}
C_1=P^aP_a+cZ^{ab}M_{ba}+2a^2M^{ab}M_{ab}\mathrel{\mathop=^{\rm def}}
X_kH_1^{kl}X_l,
\end{eqnarray}
\ba\label{2.3}
C_2=c^2Z^{ab}Z_{ab}+8a^2(cZ^{ab}M_{ba}+2a^2M^{ab}M_{ab})\mathrel{\mathop=^{\rm def}}
X_kH_2^{kl}X_l,
\ea
\ba\label{2.4}
C_3=\e^{abcd}[c^2Z_{ab}Z_{cd}+8a^2(cZ_{ba}M_{cd}+2a^2M_{ab}M_{cd})].
\ea
Note that in the limit $a\to0$ the algebra (\ref{2.1}) tend to the algebra
(\ref{1.2}) and the quadratic Casimir operators (\ref{2.2}), (\ref{2.3}) and
(\ref{2.4}) are turned into (\ref{1.3}), (\ref{1.4}) and (\ref{1.5}),
respectively.

The symmetric tensor
\ba\label{2.5}
H^{kl}=sH_1^{kl}+tH_2^{kl}=H^{lk}
\ea
with arbitrary constants $s$ and $t$ is invariant with respect to the adjoint
representation
\ba
H^{kl}=H^{mn}{U_m}^k{U_n}^l.\no
\ea
Conversely, if we demand the invariance with respect to the adjoint representation
of the second rank contravariant symmetric tensor, then we come to the structure
(\ref{2.5}) (see also the relation (32) in \cite{dss}).

The semi-simple algebra (\ref{2.1})
\ba
[X_k,X_l]={f_{kl}}^mX_m\no
\ea
has the non-degenerate Cartan-Killing metric tensor
\ba
g_{kl}={f_{km}}^n{f_{ln}}^m,\no
\ea
which is invariant with respect to the co-adjoint representation
\ba
g_{kl}={U_k}^m{U_l}^ng_{mn}.\no
\ea
With the help of the inverse metric tensor $g^{kl}$: $g^{kl}g_{lm}=\d_m^k$ we 
can
construct the quadratic Casimir operator which, as it turned out, has the
following expression in terms of the quadratic Casimir operators (\ref{2.2})
and (\ref{2.3}):
\ba\label{2.6}
X_kg^{kl}X_l={1\over8a^2(D-1)}\left[C_1+{3-2D\over8a^2(D-2)}C_2\right],
\ea
that corresponds to the particular choice of the constants $s$ and $t$ in
(\ref{2.5}).

The extended Poincar\'e algebra (\ref{2.1}) can be rewritten in the  
form:
\ba\label{2.7}
[N_{ab},N_{cd}]=(g_{ad}N_{bc}+g_{bc}N_{ad})-(c\leftrightarrow d),
\ea
\ba\label{2.8}
[L_{AB},L_{CD}]=(g_{AD}L_{BC}+g_{BC}L_{AD})-(C\leftrightarrow D),
\ea
\ba\label{2.9}
[N_{ab},L_{CD}]=0,
\ea
where the metric tensor $g_{AB}$ has the following nonzero components:
\ba
g_{AB}=\{g_{ab}, g_{D+1D+1}=-1\}.
\ea
The generators
\ba\label{2.10}
N_{ab}=M_{ab}-{c\over4a^2}Z_{ab}
\ea
form the Lorentz algebra $so(D-1, 1)$ and the generators
\ba\label{2.11}
L_{AB}=\{L_{ab}={c\over4a^2}Z_{ab}, L_{aD+1}=-L_{D+1a}={1\over2a}P_a, 
L_{D+1D+1}=0\}
\ea
form the algebra $so(D-1,2)$\footnote{Note that in the case $D=4$ we obtain the
anti-de Sitter algebra $so(3,2)$.}. The algebra (\ref{2.7})-(\ref{2.9}) is a 
direct sum~$so(D-1,1)\oplus so(D-1,2)$.

The quadratic Casimir operators $N_{ab}N^{ab}$, $L_{AB}L^{AB}$ and 
$\e^{abcd}N_{ab}N_{cd}$ of the algebra (\ref{2.7})-(\ref{2.9}) are expressed
in terms of the operators $C_1$ (\ref{2.2}), $C_2$ (\ref{2.3}) and 
$C_3$ (\ref{2.4}) in the following way:
\ba\label{2.12}
N_{ab}N^{ab}-L_{AB}L^{AB}={1\over2a^2}C_1,
\ea
\ba\label{2.13}
N_{ab}N^{ab}={1\over16a^4}C_2,
\ea
\ba\label{2.14}
\e^{abcd}N_{ab}N_{cd}={1\over16a^4}C_3.
\ea

\sect{Supersymmetric generalization}

In the case $D=4$ dimensions the extended Poincar\'e algebra (\ref{2.1}) admits
the following supersymmetric generalization:
\begin{eqnarray}
\{Q_\k,Q_\l\}=-d\left[{2a\over c}(\g^aC)_{\k\l}P_a
+(\sigma^{ab}C)_{\k\l}Z_{ab}\right],\nonumber
\end{eqnarray}
\begin{eqnarray}
[M_{ab},Q_\k]=-(\sigma_{ab}Q)_\k,\nonumber
\end{eqnarray}
\begin{eqnarray}
[P_a,Q_\k]=a(\g_aQ)_\k,\nonumber
\end{eqnarray}
\begin{eqnarray}\label{3.1}
[Z_{ab},Q_\k]=-{4a^2\over c}(\s_{ab}Q)_\k,
\end{eqnarray}
where $Q_\k$ are the super-translation generators.

Under such a
generalization the Casimir operator (\ref{2.2}) is modified by adding a term
quadratic in the super-translation generators
\begin{eqnarray}\label{3.2}
{\tilde C}_1=P^aP_a+cZ^{ab}M_{ba}+2a^2M^{ab}M_{ab}
&-&{c\over2d}Q_\k(C^{-1})^{\k\l}Q_\l\cr\nn&\stackrel{\rm def}{=}&
X_KH_1^{KL}X_L,
\end{eqnarray}
whereas the form of the rest quadratic Casimir operators (\ref{2.3}) and
(\ref{2.4}) is not changed. In (\ref{3.2}) $X_K=\{P_a, Z_{ab}, M_{ab}, Q_\k\}$ 
is a set of the generators for the also semi-simple extended superalgebra 
(\ref{2.1}), (\ref{3.1}).

The tensor
\ba\label{3.3}
H^{KL}=vH_1^{KL}+wH_2^{KL}=(-1)^{p_Kp_L+p_K+p_L}H^{LK}
\ea
is invariant with respect to the adjoint representation
\ba
H^{KL}=(-1)^{(p_K+p_M)(p_L+1)}H^{MN}{U_N}^L{U_M}^K,\no
\ea
where $p_K=p(K)$ is a Grassmann parity of the quantity $K$. 
In (\ref{3.3}) $v$ and $w$ are arbitrary constants and nonzero elements of the
matrix $H_2^{KL}$ equal to the elements of the matrix $H_2^{kl}$ followed from
(\ref{2.3}). Again, by demanding the invariance with respect to the adjoint
representation of the second rank contravariant tensor
$H^{KL}=(-1)^{p_Kp_L+p_K+p_L}H^{LK}$, we come to the structure (\ref{3.3})
(see also the relation (32) in \cite{dss}).

The semi-simple Lie superalgebra (\ref{2.1}) (\ref{3.1}) has the
non-degenerate Cartan-Killing metric tensor $G_{KL}$ (see the relation (A.5)
in the Appendix A) which is invariant with respect to the co-adjoint
representation
\ba
G_{KL}=(-1)^{p_K(p_L+p_N)}{U_L}^N{U_K}^MG_{MN}.\no
\ea
With the use of the inverse metric tensor $G^{KL}$
\ba
G^{KL}G_{LM}=\d_M^K\no
\ea
we can construct the quadratic Casimir operator (see the relation (A.8) in
the Appendix A) which takes the following expression in terms of the
Casimir operators (\ref{2.3}) and (\ref{3.2}):
\ba\label{3.4}
X_KG^{KL}X_L={1\over20a^2}\left({\tilde C}_1-{9\over32a^2}C_2\right),
\ea
that meets the particular choice of the constants $v$ and $w$ in (\ref{3.3}).

In the $D=4$ case the extended superalgebra (\ref{2.1}), (\ref{3.1}) can be
rewritten in the form of the relations (\ref{2.7})-(\ref{2.9}) and the 
following ones:
\ba\label{3.5}
\{Q_\k,Q_\l\}=-{4a^2d\over c}(\S^{AB}C)_{\k\l}L_{AB},
\ea
\ba\label{3.6}
[L_{AB},Q_\k]=-(\S_{AB}Q)_\k,
\ea
\ba\label{3.7}
[N_{ab},Q_\k]=0,
\ea
where
\ba
\S_{AB}={1\over4}[\G_A,\G_B],\quad\G_A=\{\g_a\g_5,\g_5\},\no
\ea
\ba
\{\g_a,\g_b\}=2g_{ab},\quad g_{ab}=diag(-1,1,1,1),\no
\ea
\ba
\g_5=\g_0\g_1\g_2\g_3.\no
\ea
The generators $N_{ab}$ (\ref{2.10}) form the Lorentz algebra $so(3,1)$ and 
the generators $L_{AB}$ (\ref{2.11}), $Q_\k$ form the orthosymplectic algebra
$osp(1,4)$. We see that superalgebra (\ref{2.7})-(\ref{2.9}), (\ref{3.5})-
(\ref{3.7}) is a direct sum $so(3,1)\oplus osp(1,4)$.

In this case the Casimir operator (\ref{2.12}) is modified by adding a term
quadratic in the super-translation generators
\ba
N_{ab}N^{ab}-L_{AB}L^{AB}-{c\over4a^2d}Q_\k(C^{-1})^{\k\l}Q_\l=
{1\over2a^2}{\tilde C}_1,\no
\ea
while the form of the quadratic Casimir operators (\ref{2.13}) and (\ref{2.14})
is not changed.

\sect{Conclusion}

Thus, we proposed the semi-simple second rank tensor extension of the
Poincar\'e algebra in the arbitrary dimensions $D$ and super Poincar\'e
algebra in the $D=4$ dimensions. It is very important, since under
construction of the models it is more convenient to deal with the
non-degenerate space-time symmetry. We also constructed the quadratic Casimir
operators for the semi-simple extended Poincar\'e and super Poincar\'e 
algebras.

It is interesting to develop the models based on these extended algebras. The
work in this direction is in progress.

\section*{Acknowledgments}

We are grateful to J.A. de Azcarraga for the valuable remark.
One of the authors (V.A.S.) thanks the administration of the Office of 
Associate and Federation Schemes of the Abdus Salam ICTP for the kind 
hospitality at Trieste where this work has been completed.
The research of V.A.S. was partially supported by the Ukrainian National
Academy of Science and Russian Fund of Fundamental Research, Grant No
38/50-2008.

\appendix
\sect{Appendix: Properties of Lie superalgebra}

Permutation relations for the generators $X_K$ of Lie superalgebra are
\ba\label{A.1}
\left[X_K,X_L\right\}\stackrel{\rm def}{=}
X_KX_L-(-1)^{p_Kp_L}X_LX_K={f_{KL}}^MX_M.
\ea
Structure constants ${f_{KL}}^M$ have the Grassmann parity
\ba\label{A.2}
p\left({f_{KL}}^M\right)=p_K+p_L+p_M=0\pmod2,
\ea
following symmetry property:
\ba\label{A.3}
{f_{KL}}^M=-(-1)^{p_Kp_L}{f_{LK}}^M
\ea
and obey the Jacobi identities
\ba\label{A.4}
\sum_{(KLM)}(-1)^{p_Kp_M}{f_{KN}}^P{f_{LM}}^N=0,
\ea
where the symbol $(KLM)$ means a cyclic permutation of the quantities
$K$, $L$ and $M$.
In the relations (\ref{A.1})-(\ref{A.4}) an every index $K$ takes either a
Grassmann-even value $k$ $(p_k=0)$ or a Grassmann-odd one $\k$ $(p_\k=1)$.
The relations (\ref{A.1}) have the following components:
\ba
[X_k,X_l]={f_{kl}}^mX_m,\no
\ea
\ba
\{X_\k,X_\l\}={f_{\k\l}}^mX_m,\no
\ea
\ba
[X_k,X_\l]={f_{k\l}}^\m X_\m.\no
\ea

The Lie superalgebra possesses the Cartan-Killing metric tensor
\ba\label{A.5}
G_{KL}=(-1)^{p_N}{f_{KM}}^N{f_{LN}}^M=(-1)^{p_Kp_L}G_{LK}
=(-1)^{p_K}G_{LK}=(-1)^{p_L}G_{LK},
\ea
which components are
\ba
G_{kl}={f_{km}}^n{f_{ln}}^m-{f_{k\m}}^\n {f_{l\n}}^\m,\no
\ea
\ba
G_{\k\l}={f_{\k\m}}^m{f_{\l m}}^\m-{f_{\k m}}^\m{f_{\l\m}}^m,\no
\ea
\ba
G_{k\l}=0.\no
\ea

As a consequence of the relations (\ref{A.3}) and (\ref{A.4}) the tensor with
low indices
\ba\label{A.6}
f_{KLM}={f_{KL}}^NG_{NM}
\ea
has the following symmetry properties:
\ba\label{A.7}
f_{KLM}=-(-1)^{p_Kp_L}f_{LKM}=-(-1)^{p_Lp_M}f_{KML}.
\ea

For a semi-simple Lie superalgebra the Cartan-Killing metric tensor is 
non-degenerate and therefore there exists an inverse tensor $G^{KL}$
\ba
G_{KL}G^{LM}=\d_K^M.\no
\ea
In this case, as a result of the symmetry properties (\ref{A.7}), the
quantity
\ba\label{A.8}
X_KG^{KL}X_L
\ea
is a Casimir operator
\ba
[X_KG^{KL}X_L,X_M]=0.\no
\ea

\end{document}